\documentclass[twocolumn,showpacs,superscriptaddress,preprintnumbers,amsmath,amssymb]{revtex4}
\usepackage{graphicx}
\usepackage{dcolumn}
\usepackage{bm}

\begin{document}

\preprint{}

\title{Charged particle (pseudo-)rapidity distributions in $\rm{p}$+$\bar{\rm{p}}$/$\rm{p}$+$\rm{p}$ and Pb+Pb/Au+Au collisions from SPS to LHC energies from UrQMD}

\author{Michael Mitrovski}
 \affiliation{Frankfurt Institute for Advanced Studies (FIAS),
Ruth-Moufang-Str.~1, D-60438 Frankfurt am Main,
Germany}
\affiliation{Institut f\"ur Theoretische Physik, Goethe-Universit\"at, Max-von-Laue-Str.~1, 
D-60438 Frankfurt am Main, Germany}

\author{Tim Schuster}
 \affiliation{Frankfurt Institute for Advanced Studies (FIAS),
Ruth-Moufang-Str.~1, D-60438 Frankfurt am Main,
Germany}

\author{Gunnar Gr\"af}
\affiliation{Institut f\"ur Theoretische Physik, Goethe-Universit\"at, Max-von-Laue-Str.~1,
 D-60438 Frankfurt am Main, Germany}

\author{Hannah Petersen}
\affiliation{Frankfurt Institute for Advanced Studies (FIAS),
Ruth-Moufang-Str.~1, D-60438 Frankfurt am Main,
Germany}
\affiliation{Institut f\"ur Theoretische Physik, Goethe-Universit\"at, Max-von-Laue-Str.~1, 
D-60438 Frankfurt am Main, Germany}

\author{Marcus Bleicher}
\affiliation{Institut f\"ur Theoretische Physik, Goethe-Universit\"at, Max-von-Laue-Str.~1,
 D-60438 Frankfurt am Main, Germany}

\date{\today}

\begin{abstract}
We present results for final state charged particle (pseudo-)rapidity distributions in $\rm{p}$+$\bar{\rm{p}}$/$\rm{p}$+$\rm{p}$ and Pb+Pb/Au+Au at ultra high energies (17.3 GeV $\leq$ $\sqrt{s_{\rm{NN}}}$ $\leq$ 14 TeV) from the Ultra-relativistic Quantum Molecular Dynamics (UrQMD-v2.3) model. In addition, excitation functions of produced charged particle multiplicities ($N_{\rm{ch}}$) and pseudorapidity spectra are investigated up to LHC energies. Good agreement is observed between UrQMD and measured pseudorapidity distributions of charged particles up to the highest Tevatron and Sp$\bar{\rm{p}}$S energies.
\end{abstract}

\pacs{}

\maketitle

High energy nucleon-nucleon and nucleus-nucleus collisions are an excellent tool to study nuclear matter under extreme conditions of temperature and density. First day observable is the abundance of charged particles in elementary (anti-)proton-proton collisions and in heavy-ion collisions. This allows for a first exploration of parton densities in the early stage and provides stringent limits for nearly all available theoretical models. It directly reflects how much of the initial beam energy can be converted to new particles and it is therefore directly linked to the stopping mechanism of the initial protons and nucleons. Thus, the particle multiplicity contains information about the entropy of the system and the gluon density in the first stage of the collision. In nucleus-nucleus collisions more particles are produced compared to nucleon-nucleon collisions.  By scaling the produced particle multiplicity in Pb+Pb/Au+Au collisions by $N_{\rm{part}}$ (the number of participating nucleons) it can be tested whether nucleus-nucleus collisions are just a sum of nucleon-nucleon collisions or if a more collective type of physics is taking place. The RMS-width of the charged particle pseudorapidity distribution gives information about the longitudinal expansion of the system. Starting from a model benchmark in comparison to data from SPS, RHIC and Tevatron, we proceed to a prediction for the charged particle density expected at LHC energies. 

The present letter uses the microscopic transport model UrQMD (Ultra-relativistic Quantum Molecular Dynamics) in version 2.3. This new version includes (besides other changes) a coupling to PYTHIA to allow for the treatment of hard pQCD interactions~\cite{Bleicher:1999xi,Bass:1998ca}. Further detailed explanations about the changes can be found in~\cite{Petersen:2008kb}. Therefore just a brief introduction to UrQMD will be given in this paper. UrQMD is a microscopic many body approach to p+p, p+A and A+A interactions at relativistic energies and is based on the covariant propagation of color strings, constituent quarks and diquarks accompanied by mesonic and baryonic degrees of freedom. Furthermore it includes rescattering of particles, the excitation and fragmentation of color strings and the formation and decay of hadronic resonances. Moving to higher energies more sub-hadronic degrees of freedom are available and the treatment of these is of prime importance. In the current version of UrQMD this is taken into account via the introduction of a formation time for hadrons produced in the fragmentation of strings~\cite{Andersson:1986gw,NilssonAlmqvist:1986rx,Sjostrand:1993yb} and by hard (pQCD) scattering via an embedding of the PYTHIA model. The leading hadrons of the fragmenting strings contain the valence-quarks of the original excited hadron. In UrQMD they are allowed to interact even during their formation time, with a reduced cross section where the reduction factor is defined by the additive quark model, thus accounting for the original valence quarks contained in that hadron~\cite{Bleicher:1999xi,Bass:1998ca}. Those leading hadrons therefore represent a simplified picture of the leading (di)quarks of the fragmenting string. Newly produced (di)quarks do, in the present model, not interact until they have coalesced into hadrons Ð however, they contribute to the energy density of the system. A more advanced treatment of the partonic degrees of freedom during the formation time ought to include soft and hard parton scattering~\cite{Geiger:1992ac,Geiger:1994he,Molnar:2001ux,Lin:2004en,Xu:2004mz} and the explicit time-dependence of the color interaction between the expanding quantum wave-packets~\cite{Gerland:1998bz}. However, such an improved treatment of the internal hadron dynamics has not been implemented for light quarks into the present model. 

The UrQMD model has been applied successfully to explore heavy ion reactions from BNL-AGS energies ($E_{\rm{lab}}$ = 1$A$$-$10$A$ GeV), over CERN-SPS energies ($E_{\rm{lab}}$ = 20$A$$-$160$A$ GeV) up to the full BNL-RHIC energy ($\sqrt{s_{\rm{NN}}}$ = 200 GeV). This includes detailed studies of thermalization~\cite{Bravina:1998pi,Bravina:1999dh}, particle abundances and spectra~\cite{Bass:1997xw,Bleicher:1998wv}, strangeness production~\cite{Soff:1999et}, photonic and leptonic probes~\cite{Spieles:1997ih}, J/$\Psi$s~\cite{Spieles:1999kp} and event-by-event fluctuations~\cite{Bleicher:1998wu,Bleicher:1998wd}.

\begin{figure}
\includegraphics[scale=0.47]{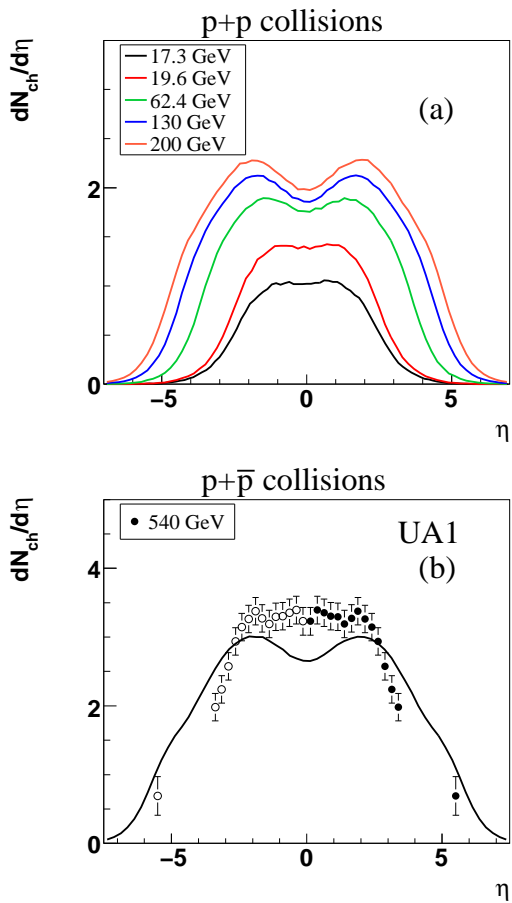}
\caption{\label{fig:fig1}(Color online) Pseudorapidity distribution of charged particles in inelastic minimum bias p+p collisions from top SPS energies to the highest RHIC energies predicted by UrQMD (a). The pseudorapidity distribution of charged particles in inelastic minimum bias p+$\bar{\rm{p}}$ collisions measured by the UA1 collaboration~\cite{Arnison:1982rm} (b). The closed symbols indicate measured points, whereas the open points are reflected with respect to mid-pseudorapidity. The solid line represents calculations from UrQMD, in inelastic minimum bias $\rm{p}$+$\bar{\rm{p}}$.}
\end{figure}
In the next Sections we set the stage for further investigations by comparing UrQMD calculations with measurements performed in p+$\bar{\rm{p}}$ and Pb+Pb/Au+Au collisions from  17.3 GeV at the CERN-SPS to 1.8 TeV at Fermilab. This systematic comparison sets the foundation for the following predictions for p+p and Pb+Pb collisions at LHC energies. 

\begin{figure}
\includegraphics[scale=0.48]{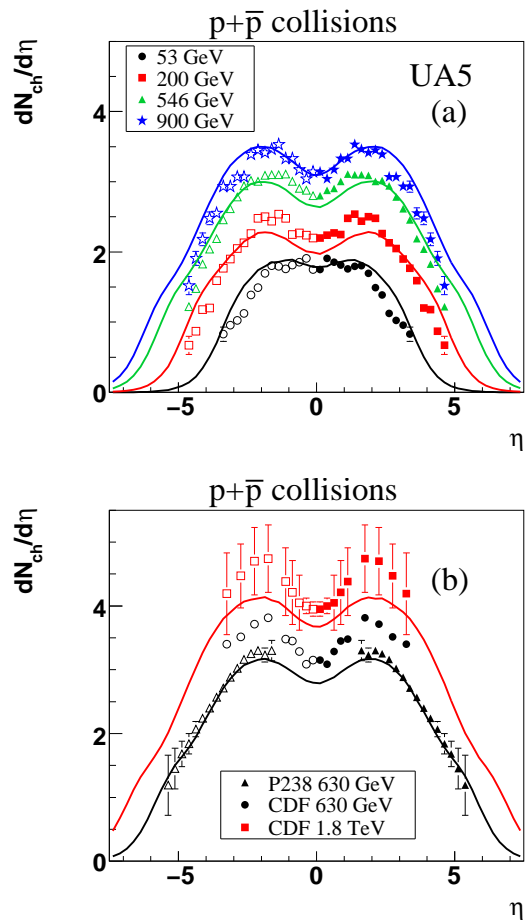}
\caption{\label{fig:fig2}(Color online) Pseudorapidity distribution of charged particles in inelastic minimum bias p+$\bar{\rm{p}}$ collisions for different energies measured by the UA5~\cite{Alner:1987wb} (a), CDF~\cite{Abe:1989td} and P238~\cite{Harr:1997sa} (b). The closed symbols indicate measured points, whereas the open points are reflected with respect to mid-pseudorapidity. The solid line represent calculations from UrQMD, in inelastic minimum bias $\rm{p}$+$\bar{\rm{p}}$.}
\end{figure}

Fig.~\ref{fig:fig1} (a) shows the d$N_{ch}$/d$\eta$ distribution ($\eta$ being the pseudorapidity) for charged particles in inelastic~\footnote{In UrQMD an inelastic collision is defined when at least one particle is created.} minimum bias p+p collisions from top SPS to top RHIC energies predicted from UrQMD. Fig.~\ref{fig:fig1} (b) presents measurements performed by the UA1 collaboration~\cite{Arnison:1982rm} for inelastic minimum bias p+$\bar{\rm{p}}$ collisions at 540 GeV. The closed points show the measured region in $\eta$, whereas the open points are the reflected points at $\eta$ = 0. With increasing energy the leading hadron effect becomes more visible and from the gap between the humps the strength of the stopping effect is visible. The system is becoming more transparent at higher energies which is reflected in the change of the pseudorapidity distribution from a Gaussian to a double Gaussian shape \cite{Bjorken:1982qr,Landau:1953gs}. The same structure is also visible for the charged particle pseudorapidity distribution in inelastic minimum bias p+$\bar{\rm{p}}$ collisions at $\sqrt{s}$ = 53, 200, 546 and 900 GeV measured by the UA5 collaboration~\cite{Alner:1987wb} (see Fig.~\ref{fig:fig2} (a)) and the P238~\cite{Harr:1997sa} and CDF~\cite{Abe:1989td} collaboration in inelastic minimum bias p+$\bar{\rm{p}}$ collisions at 630 GeV and 1.8 TeV collision energy (see Fig.~\ref{fig:fig2} (b)). A difference is observed between the experiments P238 and CDF at 630 GeV collision energy. At first glance it seems that a discrepancy between the measurements of UA1 and UA5 at 540 GeV and 546 GeV exists. However, in~\cite{Arnison:1982rm} the authors assure the reader that both experiments agree within the error, therefore we refrain from discussing possible reasons for the apparent discrepancies. 

The solid lines in Figs.~\ref{fig:fig1} and~\ref{fig:fig2} represent calculations from UrQMD in inelastic minimum bias $\rm{p}$+$\bar{\rm{p}}$ collisions. Unfortunately, no measurements of charged particle pseudorapidity distributions were performed for inelastic minimum bias p+p collisions at SPS (17.3 GeV) and RHIC energies to complete the overall picture (note however, that pion distributions at SPS and RHIC are well described by the present model~\cite{Petersen:2008kb}). Comparing UrQMD to the measurements from the UA1 (see Fig.~\ref{fig:fig1} (b)) and UA5 (see Fig.~\ref{fig:fig2} (a)) the model describes the UA1 data on a level of $\approx$ 20\% and the UA5 data within 5\% accuracy. Moving to higher energies UrQMD describes the measured peseudorapidity distribution performed by P238 (see Fig.~\ref{fig:fig2} (b)) at 630 GeV quite well. Comparing UrQMD to the measurements from CDF at 630 GeV it agrees on a level of $\approx$ 25\%. Also here, the reader should notice the difference in the measurements between P238 and CDF at 630 GeV. For the measurements at 1.8 TeV the deviation is on the level of less than 10 \%.
\begin{figure}
\includegraphics[scale=0.48]{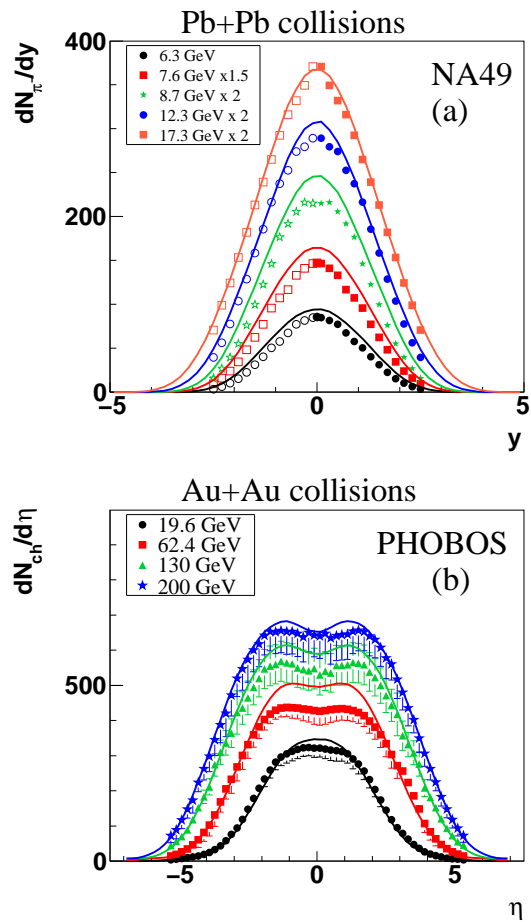}
\caption{\label{fig:fig3}(Color online) Rapidity distribution of $\pi^{-}$ in Pb+Pb collisions at SPS energies from 6.3 to 17.3 GeV (7\% most central collisions for 6.3 - 12.3 GeV, 5\% most central collisions for 17.3 GeV) measured by the NA49 collaboration~\cite{Afanasiev:2002mx,:2007fe} (a). The pseudorapidity distribution of charged particles in Au+Au collisions (6\% most central collisions, $b$ $\leq$ 3.55 fm for the data from 19.6 - 130 GeV and $b$ $\leq$ 3.65 fm for the 200 GeV dataset) at RHIC energies from 19.6 to 200 GeV performed by the PHOBOS collaboration~\cite{Abreu:2002fw,:2007we,Back:2006yw,Back:2003xk} (b). The solid line represent calculations from UrQMD ($b$ $\leq$ 3.9 fm for 7\% most central Pb+Pb collisions from 6.3 - 12.3 GeV, $b$ $\leq$ 3.4 fm for 5\% most central Pb+Pb collisions at 17.3 GeV, $b$ $\leq$ 3.6 fm for 6\% most central Au+Au collisions from 19.6 - 130 GeV and $b$ $\leq$ 3.7 fm for 6\% most central Au+Au collisions at 200 GeV).}
\end{figure}

Moving on to nucleus-nucleus reactions, Fig.~\ref{fig:fig3} shows the d$N_{\pi^{-}}$/dy and d$N_{ch}$/d$\eta$ distribution in Pb+Pb and Au+Au collisions for different experiments and energies from SPS to RHIC energies. Fig.~\ref{fig:fig3} (a) presents the dN/dy distribution of negatively charged pions measured by the NA49 collaboration~\cite{Afanasiev:2002mx,:2007fe} from 6.3 to 17.3 GeV (7\% most central collisions for 6.3 - 12.3 GeV, 5\% most central collisions for 17.3 GeV) center-of-mass energy. It is visible that UrQMD overpredicts the measurements at mid-rapidity by $\approx$ 5\% except for the ones at 17.3 GeV collision energy. Going to the higher RHIC energies (Fig.~\ref{fig:fig3} (b)) we compare to the measurements from the PHOBOS collaboration \cite{:2007we,Back:2006yw,Back:2003xk}. It is visible that the multiplicity increases with collision energy from 19.6 to 200 GeV (6\% most central collisions). Furthermore the shape of the spectra is also changing as already seen for p+p collisions due to the fact that the the colliding nuclei become increasingly transparent~\cite{Bjorken:1982qr,Landau:1953gs}. This is reflected in the UrQMD prediction where the shape of the spectra is also changing with energy. UrQMD slightly (20\%) overpredicts the measurements around mid-pseudorapidity at 62.4 GeV and 130 GeV. 
\begin{figure}
\includegraphics[scale=0.35]{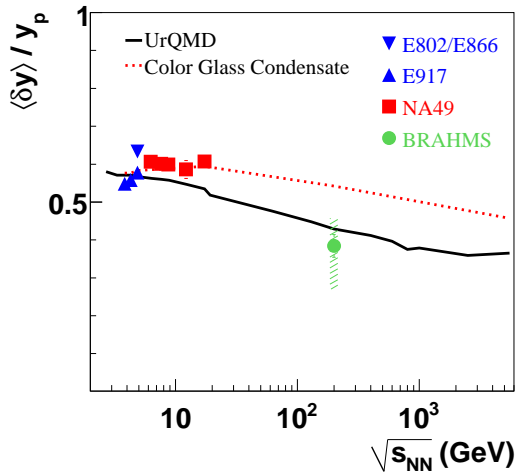}
\caption{\label{fig:fig4}(Color online) Relative rapidity shift $\left<\delta y\right>$/$y_{p}$ as a function of the center-of-mass energy in relativistic heavy ion collisions from AGS to RHIC energies \cite{Videbaek:1995mf,Back:2000ru,Appelshauser:1998yb,Bearden:2003hx,Blume:2007kw}. The black line represents the prediction made by UrQMD from low AGS to LHC energies. The dotted line represents calculations from a color glass condensate model~\cite{MehtarTani:2008qg}.}
\end{figure}

A crucial point for particle production in A+A reactions is how much of the initial longitudinal motion is transformed to particles and transverse expansion. This is best characterized by an investigation of the energy (rapidity) loss of the initial nucleons. New measurements at SPS energies (20$A$ - 80$A$ GeV) \cite{Blume:2007kw} combined with previously published results from AGS to RHIC energies \cite{Videbaek:1995mf,Back:2000ru,Appelshauser:1998yb,Bearden:2003hx} are available to test the predictions performed by the UrQMD model. Fig.~\ref{fig:fig4} depicts the energy evolution of the relative rapidity loss of the incoming nucleons in Au+Au/Pb+Pb reactions up to LHC energies. The net-baryon distribution (d$N_{B-\bar{B}}$/d$y$) is made by using the calculated rapidity spectra for $p$, $\bar{p}$, $n$, $\bar{n}$, $\Lambda$, $\Sigma^{\pm}$, $\Sigma^{0}$, $\Xi^{-}$, $\Xi^{0}$ and $\Omega^{-}$ and their anti-particles respectively. From the net-baryon distribution an average rapidity shift $\left<\delta y \right>$ can be calculated as follows:
\begin{equation}
\left<\delta y \right> = y_{p} - \frac{2}{\left<N_{\rm{part}}\right>} \int \limits_{0}^{\infty} y \frac{dN_{B-\bar{B}}}{dy} dy,
\end{equation}
where $y_{p}$ is the projectile rapidity and $\left<N_{\rm{part}}\right>$ the number of participating nucleons. It is clearly visible in the data that $\left<\delta y \right>/y_{p}$ decreases from $\approx$ 0.6 at AGS energies to 0.4 at top RHIC energies which indicates that the relative baryon stopping is slightly weaker at RHIC energies as compared to lower AGS and SPS energies. The same trend is also observed in UrQMD~\footnote{The difference to previous UrQMD versions are due to implementation of PYTHIA for p+p collisions and the resulting change in the string fragmentation function.} (black line in Fig.~\ref{fig:fig4}) where the absolute stopping follows the trend going from AGS to LHC energies. Another approach is also shown in Fig.~\ref{fig:fig4} from a color glass condensate model~\cite{MehtarTani:2008qg} (dotted line). In this model the authors are using the rapidity distribution of net protons ($p-\bar{p}$) in central heavy-ion collisions as a testing ground for saturation physics and that the valance quark parton distribution is well known at large $x$, which corresponds to the forward and backward rapidity region. 

\begin{figure}
\includegraphics[scale=0.48]{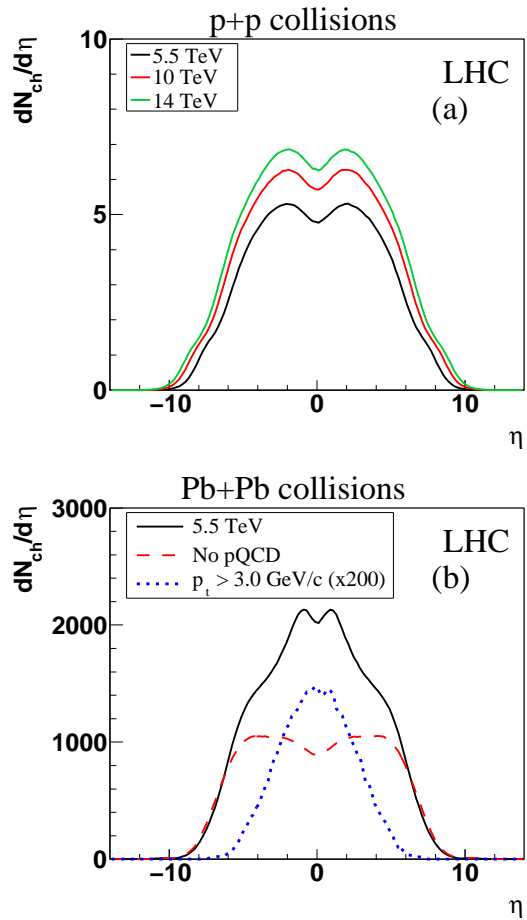}
\caption{\label{fig:fig5}(Color online) Prediction of the charged multiplicity pseudorapidity distribution for inelastic minimum bias p+p collision from $\sqrt{s_{\rm{NN}}}$ = 5.5 to 14 TeV (a) and Pb+Pb collisions (5\% most central collisions, $b$ $\leq$ 3.35 fm) at 5.5 TeV (b) collision energy from UrQMD, with PYTHIA (solid line), without pQCD contributions (PYTHIA) (dashed line) and for hard produced particles (dotted line) ($b$ $\leq$ 3.4 fm for 5\% most central Pb+Pb collisions at 5.5 TeV).}
\end{figure}

From these studies of the energy deposition (stopping) and particle production, we conclude that UrQMD has a valid basis for further extrapolations in energy and allows us to make predictions for LHC energies. 

The predictions for the charged particle pseudorapidity distributions at LHC energies are shown in Fig~\ref{fig:fig5} (a) for inelastic minimum bias p+p collisions at 5.5, 10 and 14 TeV and for the 5\% most central ($\left<N_{\rm{part}}\right>$ = 383) Pb+Pb collisions at 5.5 TeV (b) (solid line). 

\begin{figure}
\includegraphics[scale=0.33]{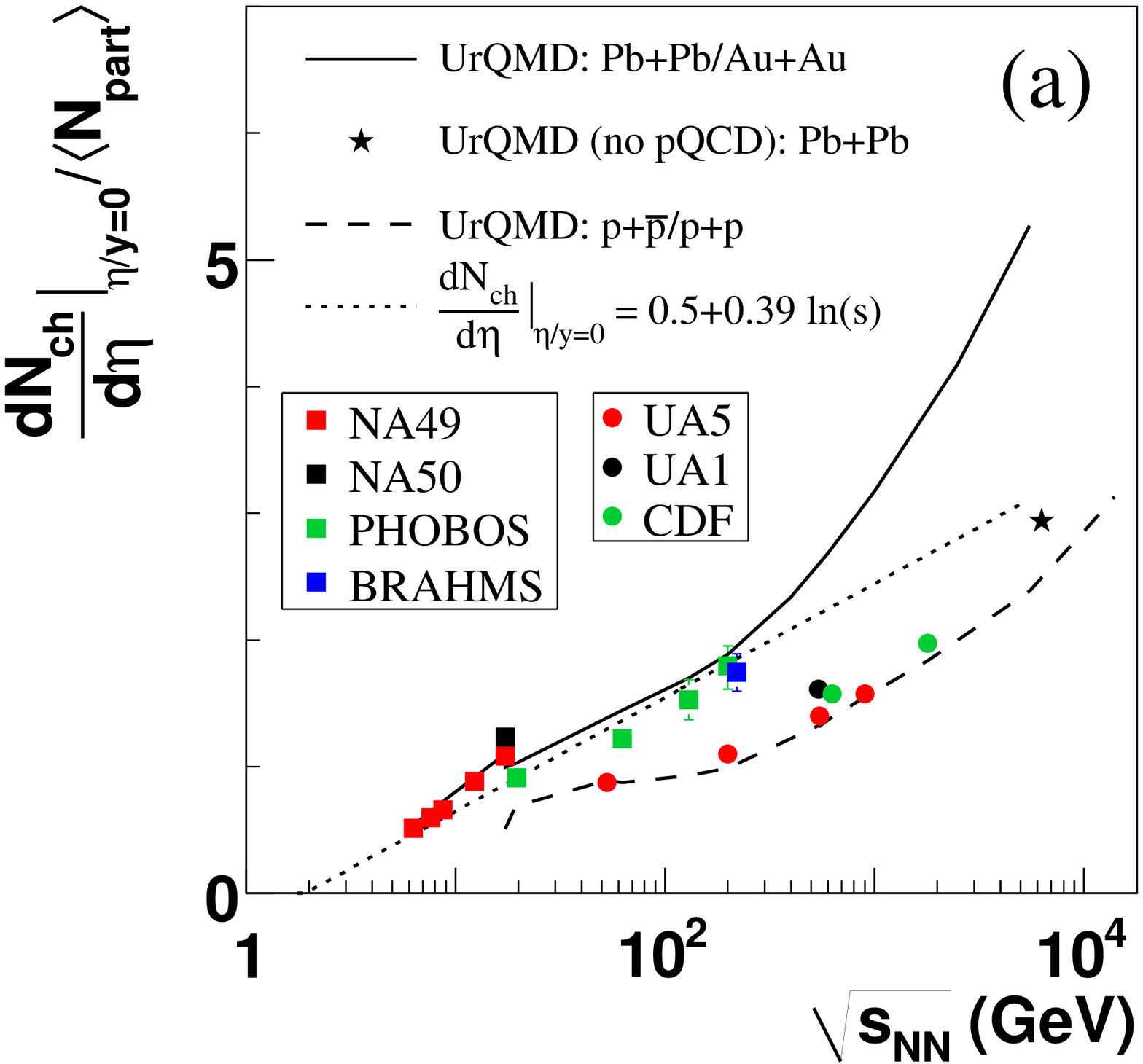}
\includegraphics[scale=0.33]{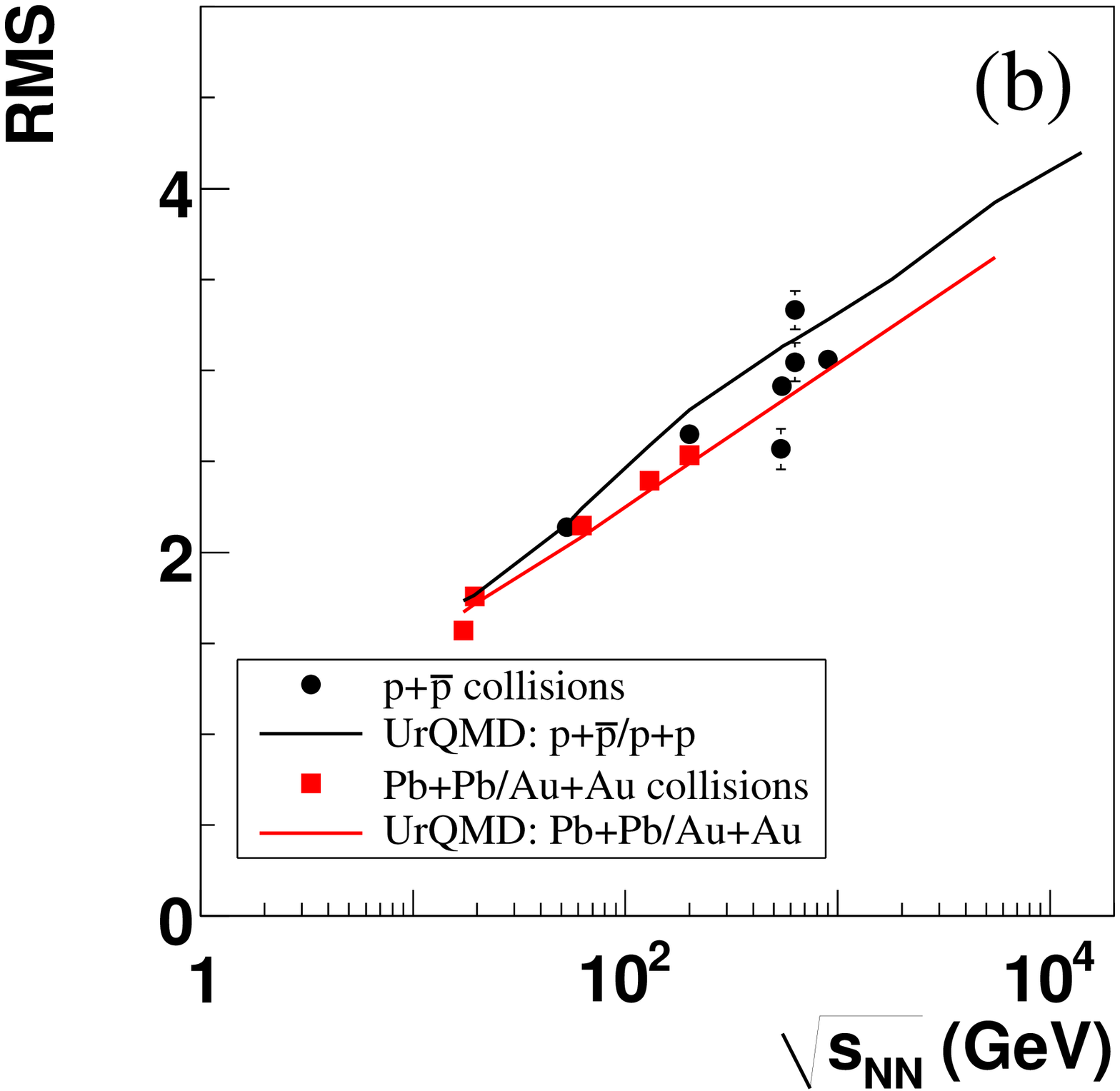}
\caption{\label{fig:fig6}(Color online) The energy dependence of the number of charged particles ($\frac{dN_{\rm{ch}}}{d\eta}$) at mid-pseudorapidity for p+$\bar{\rm{p}}$ (circles) and Pb+Pb/Au+Au (squares) collisions divided by $N_{\rm{part}}$ (a). RMS width of the pseudorapidity rapidity distributions as a function of the center-of-mass energy (b). The black solid line represents calculations from UrQMD for p+$\bar{\rm{p}}$/p+p and the red solid line for Pb+Pb/Au+Au collisions respectively.}
\end{figure}

There are two complementary production mechanisms at LHC energies: hard parton-parton scattering and soft processes. Particles produced in hard scatterings are usually created in primary collisions and are centered in a narrow region around mid-pseudorapidity (seen in dotted line in Fig.~\ref{fig:fig5} (b)), whereas soft produced particles are distributed over the full pseudorapidity range (see dashed line in Fig.~\ref{fig:fig5} (b)). At LHC energies both mechanisms play an important role so that the pseudorapidiy distribution of charged particles shown in Fig~\ref{fig:fig5} (b) (solid line) is the sum of both processes. 

\begin{figure}
\includegraphics[scale=0.38]{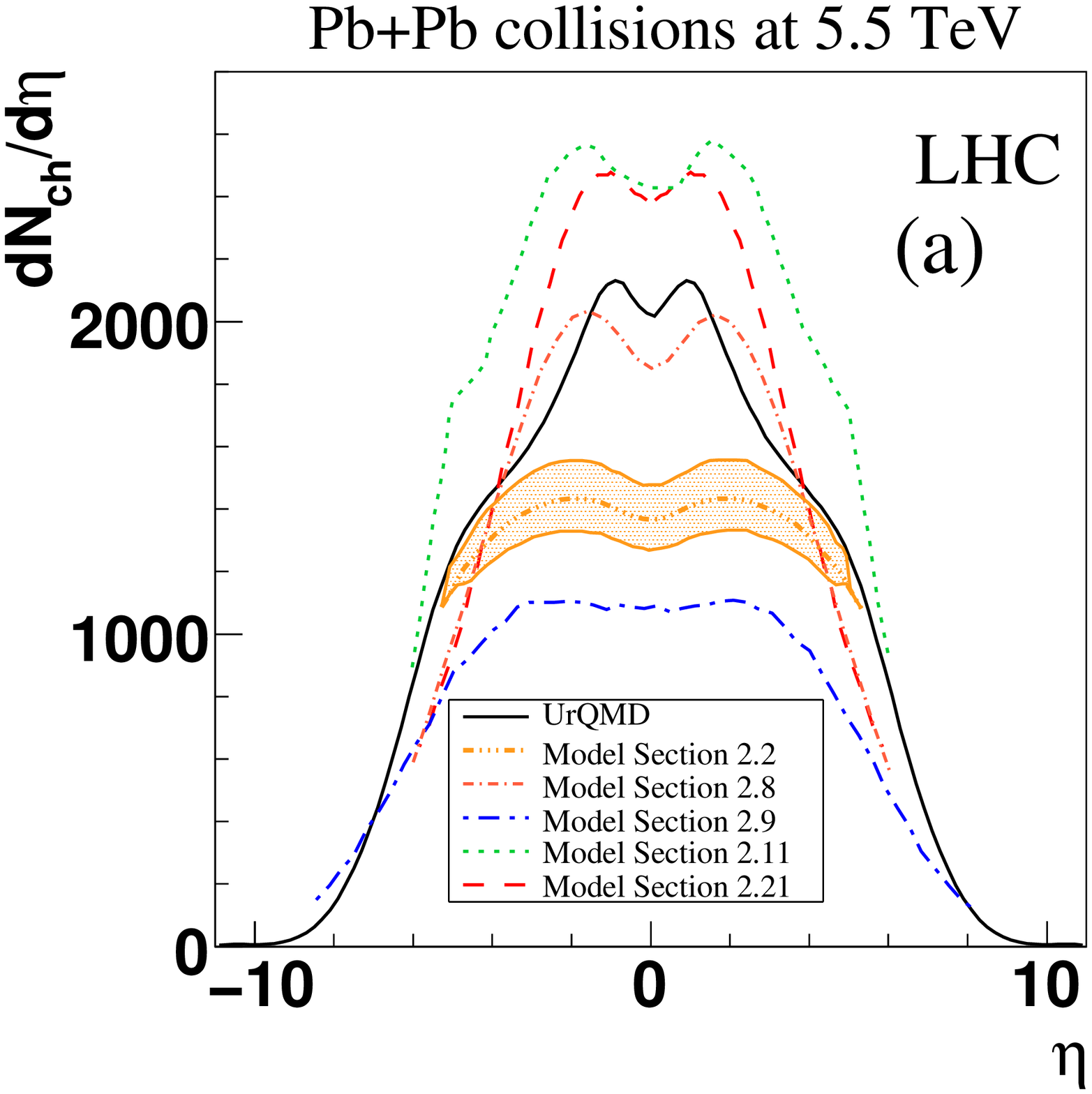}
\includegraphics[scale=0.38]{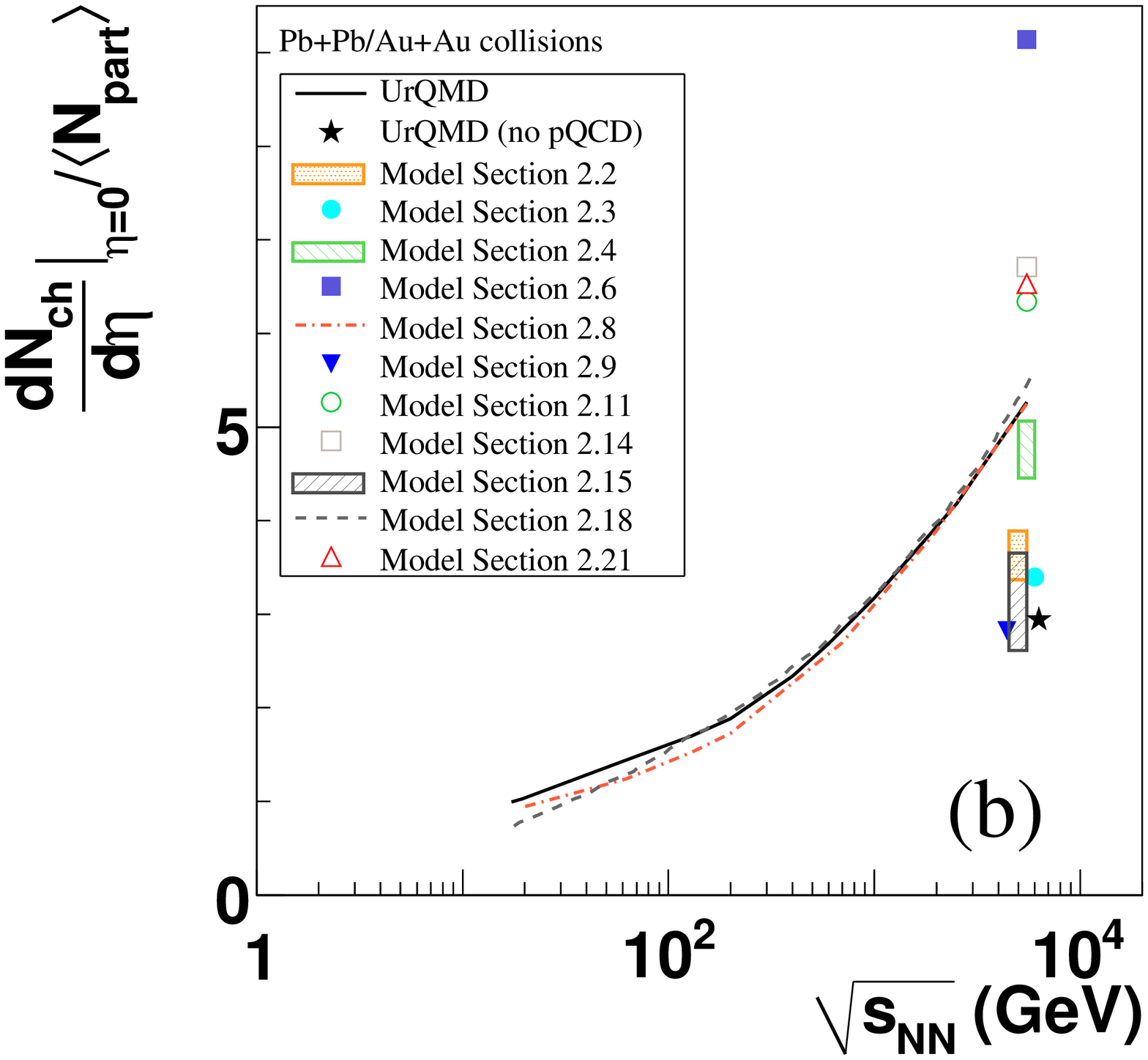}
\caption{\label{fig:fig7}(Color online) Comparison of the predicted pseudorapidity distribution of charged particles (a) and the charged particle multiplicity at mid-pseudorapidity (b) from UrQMD and predictions from various other models~\cite{Abreu:2007kv}.}
\end{figure}

Fig.~\ref{fig:fig6} (a) shows the measured number of charged particles at mid-pseudorapidity ($\frac{dN_{\rm{ch}}}{d\eta}$$\mid_{\eta/y=0}$) as a function of $\sqrt{s_{NN}}$ for p+$\bar{\rm{p}}$ (circles)~\cite{Arnison:1982rm,Alner:1987wb,Abe:1989td} and Pb+Pb/Au+Au (squares)~\cite{Afanasiev:2002mx,:2007fe,Abreu:2002fw,:2007we,Back:2003xk,Back:2006yw,Bearden:2001qq} collisions~\footnote{Note that the number of charged particles for NA49 is calculated by adding the midrapidity yields of $\pi^{-}$, $\pi^{+}$, $K^{-}$ and $K^{+}$.}. It is clearly visible that in A+A collisions $N_{\rm{ch}}$ scales linearly with the center-of-mass energy. The difference in scaling with $N_{\rm{part}}$ between p+$\bar{\rm{p}}$/p+p and Pb+Pb/Au+Au collisions increases with increasing center-of-mass energy. A simple approach to extrapolate the number of charged particles in Pb+Pb collisions was suggested in~\cite{Abreu:2007kv} by using a fit function ($\frac{dN_{\rm{ch}}}{d\eta}$$\mid_{\eta/y=0}$ = 0.5+0.39$\cdot$ln(s)). It is visible that the fit function and UrQMD agree until top RHIC energies. At higher energies UrQMD predicts a higher multiplicity in central Pb+Pb collisions, especially for top LHC energies as compared to the simple extrapolation. The reason for the increasing numbers of the multiplicity is the increase of hard collisions at LHC energies. When not taking hard collisions into account (see Fig.~\ref{fig:fig6} (a)) by switching off PYTHIA and just allow UrQMD to have soft particle production, UrQMD would follow the simple linear fit function. If the LHC data fall on the dotted line, hard collisions are either absent at LHC or saturation effects do effectively suppress a large part of the particle production. UrQMD not only describes the multiplicity and trend in p+$\bar{\rm{p}}$/p+p collisions (dashed line) but also in Pb+Pb/Au+Au collision (solid line). Furthermore in UrQMD, if going to LHC energies, the difference between p+p and Pb+Pb collisions becomes larger. 

The RMS-width~\footnote{RMS = $\sqrt(\eta_{0}^{2}+\sigma^{2}$)} is calculated by fitting the measured pseudorapidity distribution of charged particles from UA1, UA5, P238 and CDF experiments for $\rm{p}$+$\bar{\rm{p}}$ NA50 and PHOBOS for Pb+Pb/AuAu collisions by a double Gaussian~\footnote{Where double Gaussian means that we parametrized the pseudorapidity distribution by the sum of two Gauss distributions placed symmetrically with respect to mid-pseudorapidity and defined as follows:  dN/d$\eta$ = N ($e^{- \frac{\eta - \eta_{0}}{2\sigma^{2}}}$ + $e^{- \frac{\eta + \eta_{0}}{2\sigma^{2}}}$), where $\eta_{0}$ is the mean and $\sigma^{2}$ the variance of the distribution.} (see Fig.~\ref{fig:fig6} (b)). An increase of the RMS-width is observed for p+$\bar{\rm{p}}$ and Pb+Pb/Au+Au collisions with the center-of-mass energy. The dependence is linear for p+$\bar{\rm{p}}$ and Pb+Pb/Au+Au collisions. In the data, no difference between the RMS-width in p+$\bar{\rm{p}}$ and Pb+Pb/Au+Au is visible. UrQMD shows a slight difference between the RMS-width for p+$\bar{\rm{p}}$ and Pb+Pb/Au+Au collisions. 

To have an overall picture how the presented prediction of UrQMD compares to other approaches Fig.~\ref{fig:fig7} depicts the compiled results from other model predictions. Fig.~\ref{fig:fig7} (a) shows the predicted pseudorapidity distributions of charged particles from various models~\cite{Abreu:2007kv} in comparison to UrQMD. It is visible that all transport models (hadronic/partonic), including UrQMD, can be put together in one group by predicting a similar shape and multiplicity. The second group are saturation models which in general predict a lower multiplicity (also seen in~\cite{Armesto:2000xh}). This is also visible in Fig.~\ref{fig:fig7} (b) where the energy dependence of predicted charged particle multiplicity at mid-pseudorapidity is shown. At first glance it seems that the data would follow more the trend of a straight line but the major part of the models including UrQMD do not favour this trend (also seen in~\cite{Sarkisyan:2005rt}). 

In this paper we presented LHC predictions from the Ultra-relativistic Quantum Molecular Dynamics model (UrQMD). We started by showing that UrQMD describes the charged particle pseudorapidity spectra in p+$\bar{\rm{p}}$ as well as for Pb+Pb/Au+Au collisions up to Tevatron energies. Furthermore it also describes the energy dependence of charged particles in mid-pseudorapidity in p+$\bar{\rm{p}}$ and Pb+Pb/Au+Au collisions. The observed similar RMS-width of the charged particle pseudorapidity distribution in p+$\bar{\rm{p}}$ and Pb+Pb/Au+Au collisions can also be described by our model. At LHC we predict $dN_{ch}/d\eta \mid_{14\rm{A} \  \rm{TeV} \ p+p} \ \approx$ 6.3 and $dN_{ch}/d\eta \mid_{5.5 \ \rm{TeV} \ Pb+Pb} \ \approx$ 2000.

This work was supported by the Hessian LOEWE initiative through HIC for FAIR. We are grateful to the Center for Scientific Computing (CSC) at Frankfurt for the computing resources. H.~Petersen gratefully acknowledges financial support by the Deutsche Telekom Stiftung and support from the Helmholtz Research School on Quark Matter Studies. T.~Schuster gratefully acknowledges support from the Helmholtz Research School on Quark Matter Studies. This work was supported by GSI, BMBF and DESY. The authors would also like to thank C. Blume.

\end{document}